\documentclass[a4paper]{article}
\usepackage{amsfonts,amssymb,amsmath,amsthm,cite}
\usepackage{ucs} 
\usepackage[utf8x]{inputenc}
\usepackage{graphicx}
\usepackage[english]{babel}
\usepackage{slashed}
\usepackage{textcomp}
\usepackage{bm}
\usepackage{titlesec}
\usepackage{stackengine}

\usepackage{etoolbox}
\patchcmd{\thebibliography}{\section*}{\section}{}{}
\usepackage{titlesec}
\titleformat*{\section}{\normalsize\bf}
\usepackage{adjustbox}
\usepackage{nccmath}
\usepackage{mathtools}
\usepackage{xcolor}
\usepackage{footnote}
\makesavenoteenv{tabular}
\mathtoolsset{showonlyrefs}
\evensidemargin=\oddsidemargin
\begin{document}
\vspace{10mm}
	
	\begin{center}
		\large{\textbf{Three-loop functions for a quartic model with a cutoff}}
	\end{center}
	\vspace{2mm}
	\begin{center}
		\textbf{A. V. Ivanov${}^{\dagger}$}\,\,\,\,\,\,\textbf{V. A. Nikiforov${}^{\star}$}
	\end{center}
	\begin{center}
		${}^{\dagger}$St. Petersburg Department 
of Steklov Mathematical Institute 
of Russian Academy of Sciences,\\ 27 Fontanka, St. Petersburg, 191023, Russia
	\end{center}
\begin{center}
	E-mail: regul1@mail.ru
\end{center}
	\begin{center}
		 ${}^{\star}$National Research University Higher School of Economics,\\ 3 Kantemirovskaya St., Saint Petersburg, 194100, Russia
	\end{center}
	\begin{center}
		E-mail: nikiforoff.voff@yandex.ru
	\end{center}

	\vspace{15mm}
	
\noindent\textbf{Abstract.} This paper presents numerical values for auxiliary integrals and coefficients of the beta function in the three-loop approximation for a four-dimensional model with a quartic interaction, using a special type of regularization function. The values are compared to previously obtained results.

	\vspace{2mm}

\newpage	
\section{Introduction}
\label{v-s-1}
Multiloop calculations in the framework of perturbative quantum field theory \cite{1,2} are closely related to numerical analysis methods, since the resulting integrals cannot always be expressed in terms of elementary functions. In this context, the most popular is a four-dimensional model with a quartic interaction, whose classical action is described by the functional
\begin{equation*}
\int_{\mathbb{R}^4}\mathrm{d}^4x\,\bigg[
\frac{1}{2}\big(\partial_{x_\mu}\phi(x)\big)\big(\partial_{x^\mu}\phi(x)\big)+
\frac{m^2}{2}\phi^2(x)+\frac{\lambda}{4!}\phi^4(x)\bigg].
\end{equation*}
This model is important in the theory of critical phenomena, see \cite{29-4,29-3}, as it has a number of nontrivial properties and at the same time allows many calculations to be performed explicitly. One of these properties is renormalizability, which allows, after the introduction of a regularization, to eliminate singular contributions by multiplying parameters and fields by renormalization constants
\begin{equation*}
\phi\to\sqrt{Z_2}\phi,\,\,\,m^2\to ZZ_2^{-1}m^2,\,\,\,\lambda\to Z_4^{\phantom{1}}Z_2^{-2}\lambda\equiv\lambda_0,
\end{equation*} 
which are functions of the regularizing parameter. For example, in the case of dimensional regularization \cite{19,555}, such a parameter is a small deviation $\varepsilon$ from the basic dimension, and in the case of cutoff, a large momentum $\Lambda$. In both situations, an auxiliary dimensionless momentum $\mu$ appears in the constants, on which the renormalized coupling constant $\lambda=\lambda(\mu)$ depends. Note that for convenience in the dimensional regularization, the auxiliary multiplier $\mu^{\varepsilon}$ is formally assigned to $Z_4$. As is known, in the theory of critical behavior, an important role is played by the $\beta$-function and the anomalous dimensions $\gamma_{\phi}$ and $\gamma_{\tau}$, which according to \cite{29-3} are determined by the equations
\begin{align*}
\beta(\lambda)&=\mu\frac{\mathrm{d}\lambda}{\mathrm{d}\mu}=
\beta_1\hat{\lambda}^2+
\beta_2\hat{\lambda}^3+
\beta_3\hat{\lambda}^4+\ldots,\\
\gamma_{\phi}(\lambda)&=\mu\frac{\mathrm{d}\ln(Z_2)}{2\mathrm{d}\mu}=
\gamma_{\phi,1}\hat{\lambda}+
\gamma_{\phi,2}\hat{\lambda}^2+
\gamma_{\phi,3}\hat{\lambda}^3+\ldots,
\\
\gamma_{\tau}(\lambda)&=\mu\frac{\mathrm{d}\ln(Z/Z_2)}{\mathrm{d}\mu}=
\gamma_{\tau,1}\hat{\lambda}+
\gamma_{\tau,2}\hat{\lambda}^2+
\gamma_{\tau,3}\hat{\lambda}^3+\ldots.
\end{align*}
Here $\hat{\lambda}=\lambda/(16\pi^2)$. Note that in the case of a cutoff regularization, $Z$ must be replaced by the logarithmic part, without power singularities. It is known \cite{105} that the coefficients $\{\beta_1,\beta_2,\gamma_{\phi,1},\gamma_{\tau,1}\}$ are independent of the subtraction scheme. The remaining coefficients depend on the choice and are presented within the framework of this work using the $\mathrm{MS}$-scheme. So, turning to the well-known two-loop coefficients, see \cite{v-2,v-3,v-15}, we can write out
\begin{equation*}
\beta_1=3,\,\,\,\beta_2=-17/3,\,\,\,
\gamma_{\phi,1}=0
,\,\,\,
\gamma_{\phi,2}=1/12
,\,\,\,
\gamma_{\tau,1}=-1
,\,\,\,
\gamma_{\tau,2}=5/6
.
\end{equation*}
At the same time, the subsequent coefficients also depend on the type of regularization, and in a significant way. For example, in the dimensional regularization in the $\mathrm{MS}$-scheme, the third coefficients are equal 
\begin{equation*}
\beta_3^{\mathrm{DR}}=12 \zeta(3)+145/8\approx32,54968\pm10^{-5},\,\,\,
\gamma_{\phi,3}^{\mathrm{DR}}=-1/16,\,\,\,
\gamma_{\tau,3}^{\mathrm{DR}}=-7/2.
\end{equation*}
The results for the $\overline{\mathrm{MS}}$-scheme can be found, for example, in \cite{v-60}.
Note that the dimensional deformation approach has progressed quite a lot. At the moment, not only the four-loop corrections \cite{v-1,v-4} are known, but also higher ones, see \cite{v-5,v-55}. In the case of another type of regularization, until recently only two-loop coefficients were known. For example, see the case of implicit regularization \cite{chi-0} or a cutoff regularization \cite{v-10,v-11,v-12} as part of studying the functional renormalization group \cite{v-13,v-14}. 

In the recent paper \cite{3}, the three-loop singular contribution was studied for a family of cutoff regularizations in the coordinate representation \cite{4}, depending on a continuous regularizing function $\mathbf{f}(\cdot)$ with a support in $[0,1]$. When examining the third coefficients, it became clear that they are finite combinations of integrals, each of which depends on the function $\mathbf{f}(\cdot)$. In the same paper, a numerical analysis was performed for the simplest case $\mathbf{f}=0$. The result for the scheme with minimal subtractions is
\begin{equation*}
\beta_3^{\mathbf{0}}\approx45,90683\pm10^{-5},\,\,\,
\gamma_{\phi,3}^{\mathbf{0}}=-1/24,\,\,\,
\gamma_{\tau,3}^{\mathbf{0}}=-127/36.
\end{equation*}
However, the considered regularization option does not satisfy the condition of applicability of the cutoff \cite{5}, which would guarantee the non-negativity of the spectrum of the deformed free Laplace operator. Therefore, it remained an open problem to conduct a numerical analysis for a case that would satisfy such a condition.

\section{Results}
\label{v-s-2}
This paper is devoted to the numerical analysis of a set of auxiliary integrals $\alpha_1(\mathbf{f})$--$\alpha_{13}(\mathbf{f})$, in which the regularizing function is chosen as follows, see \cite{5},
\begin{equation}\label{111}
\mathbf{f}(s)\to\mathbf{f}_4(s) = 3 - \frac{4s + 2}{\pi}\Big(\frac{1-s}{s}\Big)^{\frac{1}{2}} 
	+ \frac{2}{\pi}\Big(\frac{1}{s}-4\Big)\arcsin(\sqrt{s}).
\end{equation}
\begin{table}[h!]
	\centering
	\renewcommand{\arraystretch}{1.2}
	\resizebox{9cm}{!}{
		\begin{tabular}{ |l||l||l| }
			\hline&
			\multicolumn{1}{|c||}{Case $\textbf{f} = 0$.} &
			\multicolumn{1}{|c|}{Case $\textbf{f}=\textbf{f}_4$.} \\
			\hline\hline
			
			\centering $\alpha_1$ &  
			$ -0.125$   &   $-0.0625$ \\
			\hline
			
			\centering $\alpha_2$ &  
			$ \phantom{-}0.25$  &   $\phantom{-}0.61353797\pm10^{-8}$ \\
			\hline
			
			\centering $\alpha_3\times16\pi^4$ & 
			$\phantom{-}0.375$ &  $\phantom{-}0.80175006\pm10^{-8}$ \\
			\hline
			
			\centering $\alpha_4\times96\pi^4$ &  
			$\phantom{-}0.125$   &   $\phantom{-}0.55540693\pm 10^{-8}$ \\
			\hline
			\centering$\alpha_5\times16\pi^4$ &  
			$-0.04298947\pm 10^{-8}  $ & $-0.04491004\pm 10^{-8}$\\
			\hline
			\centering$\alpha_6\times3$ &  
			$\phantom{-}0.5 $& $\phantom{-}1.45765688\pm10^{-8}$ \\
			\hline
			\centering$\alpha_7\times64\pi^6$ &  
			$-0.02667863\pm 10^{-8} $ & $ -0.06175358\pm10^{-8}$\\
			\hline
			\centering$\alpha_8\times384\pi^6$ & 
			$\phantom{-}0.125 $ & $\phantom{-}0.09951213\pm 10^{-8}$\\
			\hline
			\centering$\alpha_9\times64\pi^6$ &  
			$-0.02771715\pm 10^{-8} $  & $-0.06496642\pm10^{-8}$\\
			\hline
			\centering$\alpha_{10}\times384\pi^6$ &  
			$\phantom{-}0.125 $  & $\phantom{-}1.07159869\pm 10^{-8}$\\
			\hline
			\centering$\alpha_{11}\times96\pi^4$ & 
			$\phantom{-}0.125$ & $\phantom{-}0.64541799\pm10^{-8}$\\
			\hline
			\centering$\alpha_{12}\times24\pi^2$ &  
			$\phantom{-}0.125 $ & $\phantom{-}0.31614873\pm10^{-8}$\\
			\hline
			\centering$\alpha_{13}\times4\pi^2$ &  
			$-0.01122021\pm 10^{-8} $ & $-0.01644843\pm10^{-8}$  \\
			\hline
	\end{tabular}}
	\renewcommand{\arraystretch}{1}
	\caption{Numerical values of the functionals $\alpha_1(\mathbf{f})$--$\alpha_{13}(\mathbf{f})$ from \cite{3} with $\mathbf{f}=0$ and $\mathbf{f}=\mathbf{f}_4$. If there is no error, an accurate value is provided.}
	\label{table:1}
\end{table}

\noindent Definitions for integrals are presented in Section \ref{v-s-3}, and the results are given in Table \ref{table:1}. Based on the obtained values for the case of the scheme with minimal subtractions, the third coefficients are found, which, according to \cite{3}, are determined by formulas of the form 
\begin{align*}
		\beta_3 &= 77/4 + 12 \zeta(3) - 36\alpha_1 + 17\alpha_2 + 2\alpha_6 + 18\alpha_2^2
		+(4\pi^2)^2(16\alpha_{11} - 4\alpha_{3} - 144\alpha_{4} - 144\alpha_{5}),\\
		\gamma_{\phi,3}&=-\alpha_2/2+\alpha_6/2
		,\\
		\gamma_{\tau,3}&=-47/12+2\alpha_2/3-\alpha_6+(4\pi^2)^2(4\alpha_3/3-16\alpha_{11}/3)
		,
\end{align*}
where $\zeta$ is the Riemann zeta function. After substituting the values from Table \ref{table:1}, we get
\begin{equation*}
	\beta_3^\mathbf{4}\approx 45.75371 \pm 10^{-5},\,\,\,
	\gamma_{\phi,3}^{\mathbf{4}}\approx-0.0638262\pm 10^{-7},\,\,\,
	\gamma_{\tau,3}^{\mathbf{4}}\approx-3.4982318\pm 10^{-7}.
\end{equation*}
At the same time, the relations are fulfilled for the known coefficients
\begin{equation*}
\beta_3^{\mathbf{0}}>\beta_3^\mathbf{4}>\beta_3^{\mathrm{DR}}>0.
\end{equation*}
However, as practice shows, the anomalous dimension can be compared not only to the standard renormalization constant. Denote by $z_{\Lambda,i}$ the part of $z_i$ from the constant $Z=1+\sum_{i>0} z_i$ proportional to $\Lambda^2$, then we can calculate the value of the form
\begin{equation*}
\gamma_\Lambda(\lambda)=\mu\frac{\mathrm{d}\ln(Z_\Lambda)}{\mathrm{d}\mu}=
\gamma_{\Lambda,1}\hat{\lambda}+
\gamma_{\Lambda,2}\hat{\lambda}^2+\ldots,
\,\,\,\mbox{where}\,\,\,
Z_\Lambda=1+\sum_{i>1}\frac{z_{\Lambda,i}}{z_{\Lambda,1}}.
\end{equation*}
The results of the three-loop analysis allow us to determine the first two coefficients. The first one is scheme independent and is equal to $\gamma_{\Lambda,1}=-3$. The second one depends on the procedure and in the case of using the $\mathrm{MS}$-scheme is equal to $\gamma_{\Lambda,2}=35/6$.

\section{Definitions for functionals}
\label{v-s-3}
\noindent Analytical expressions for integrals:
\begin{equation*}
\renewcommand{\arraystretch}{2}
\begin{tabular}{ll}
	${\displaystyle \alpha_1 ( \textbf{f}) = -\frac{1}{4}+ 4\pi^2h_1( y)\big|_{|y|=1},}$&${\displaystyle \alpha_7( \textbf{f}) = \int_{\mathbb{R}^{8}} \mathrm{d}^4x\mathrm{d}^4y\, \big( R_0^1(x) \big)^2 f_1^1(x+y)\big( R_0^1(y) \big)^2,}$\\
	${\displaystyle \alpha_2 ( \textbf{f}) = 2(4\pi^2)^2h_2( y)\big|_{|y|=1},}$&${\displaystyle \alpha_8( \textbf{f}) = \int_{\mathrm{B}_1} \mathrm{d}^4y\, h_2(y)\big(f_2^1(y)\big)^2,}$\\
	${\displaystyle \alpha_3 ( \textbf{f}) = \frac{1}{4(4\pi^2)^2}+4\pi^2h_3( y)\big|_{|y|=1},}$&${\displaystyle \alpha_9( \textbf{f}) = \int_{\mathbb{R}^{8}} \mathrm{d}^4x\mathrm{d}^4y\, \big( R_0^1(y) \big)^3 f_1^1(x+y) R_0^1(x),}$\\
	${\displaystyle \alpha_4 ( \textbf{f}) = \int_{\mathrm{B}_{1}}\mathrm{d}^4y\,h_2(y)f^1_2(y),}$&${\displaystyle \alpha_{10}( \textbf{f}) = \int_{\mathrm{B}_1} \mathrm{d}^4y\, h_1(y)\big(f_2^1(y)\big)^3,}$\\
	${\displaystyle \alpha_5( \textbf{f}) = \int_{\mathbb{R}^{8}} \mathrm{d}^4x\mathrm{d}^4y\,\left( R_0^1(x) \right)^2 f_1^1(x+y)R_0^1(y),}$&${\displaystyle \alpha_{11}( \textbf{f}) = \int_{\mathrm{B}_1} \mathrm{d}^4x\, h_3(y),}$\\
	${\displaystyle \alpha_6( \textbf{f}) = 2(4\pi^2)^2 \int_{\mathrm{B}_1} \mathrm{d}^4y\, \big(f_2^1(y)\big)^3 |y|^2,}$&${\displaystyle \alpha_{12}( \textbf{f}) = \int_{\mathrm{B}_1} \mathrm{d}^4x\, f_2^1(y)  h_1(y),}$\\
	\multicolumn{2}{l}{${\displaystyle \alpha_{13}( \textbf{f}) = \int_{{\mathrm{B}_1}\times{\mathrm{B}_1}}\mathrm{d}^4x\mathrm{d}^4y\,
			f_1^1(x) R_0^1(x+y) f_1^1(y)
			+ 
			\int_{\mathrm{B}_1} \mathrm{d}^4x\, f_1^1(x) h_1(x).}$}
\end{tabular}
\renewcommand{\arraystretch}{1}
\end{equation*}
\noindent Expressions for auxiliary functions:
\begin{equation*}
	\renewcommand{\arraystretch}{2}
	\begin{tabular}{l}
		${\displaystyle h_i(y) = \frac{1}{4\pi^2|y|^2}\int_{\mathrm{B}_{|y|}}\mathrm{d}^4x\,\big(f^1_2(x)\big)^i+\frac{1}{4\pi^2}\int_{\mathrm{B}_1 \backslash \mathrm{B}_{|y|}}\mathrm{d}^4x\,\big(f^1_2(x)\big)^i\frac{1}{|x|^2},}$\\
		${\displaystyle R_0^{1}(x) = \frac{\textbf{f}(|x|^2)}{4\pi^2}+\frac{1}{4\pi^2}  
			\begin{cases}
				1, & \text{$ |x| \leqslant 1 $};\\
				1/|x|^2, & \text{$ |x| > 1 $},
		\end{cases}}$\\
		${\displaystyle f_2^{1}(x) = \frac{1}{4\pi^2}\left( 1+\mathbf{f}(|x|^2)\right)\chi(|x| \leqslant 1),}$\\
		${\displaystyle f_1^{1}(x) = f_2^{1} - \frac{\chi(|x| \leqslant 1)}{4\pi^2 |x|^2}.}$
	\end{tabular}
	\renewcommand{\arraystretch}{1}
\end{equation*}

\section{Calculation methods}
\label{v-s-4}
Let us briefly describe the main stages of the calculation. First, the integrals should be simplified by switching to spherical coordinates. Let us demonstrate this using the example of $\alpha_5(\textbf{f})$, the rest are transformed similarly. To do this, consider the Fourier transform in four-dimensional space for the function $F(x)$, depending on the absolute value of $s=|x|$,
\begin{equation}
t\widetilde{F}(k) = (2\pi)^{2} \int_{\mathbb{R}_+}\mathrm{d}s\,J_{1}(ts)F(x)s^2,
\end{equation}
where $k\in\mathbb{R}^4$ and $t=|k|$. Here $J_1(\cdot)$ is a Bessel function of the first kind. Then, using the convolution transformation theorem, we get
\begin{equation}
\alpha_5( \textbf{f}) = 
8\pi^4 
\int_{\mathbb{R}_+}\mathrm{d}t\, \bigg[
\int_{\mathbb{R}_+}\mathrm{d}s\, J_{1}(ts)s^2 (R_0^1(x))^2
\int_{\mathbb{R}_+}\mathrm{d}u\, J_{1}(tu)u^2 f_1^1(y)
\int_{\mathbb{R}_+}\mathrm{d}v\, J_{1}(tv)v^2 R_0^1(z)
\bigg],
\end{equation}
where $s=|x|$, $u=|y|$, and $v=|z|$. Thus, the dimension of the integral has decreased from 8 to 4. The second step is to substitute the Fourier transforms for those functions for which they are calculated explicitly. There are several such functions and, taking into account the result from \cite{4}, they have the form
\begin{equation}
\int_{\mathbb{R}_+}\mathrm{d}s\, J_{1}(ts)s^2 R_0^1(x)\big|_{\mathbf{f}=0}=\frac{J_1(t)}{2\pi^2t^2},\,\,\,
\int_{\mathbb{R}_+}\mathrm{d}s\, J_{1}(ts)s^2 R_0^1(x)\big|_{\mathbf{f}=\mathbf{f}_4}=\frac{4J_1^2(t/2)}{\pi^2t^3},
\end{equation}
\begin{equation}
\int_{\mathbb{R}_+}\mathrm{d}s\, J_{1}(ts)s^2 f_2^1(x)\big|_{\mathbf{f}=0}=\frac{J_1(t)}{2\pi^2t^2}-\frac{J_0(t)}{4\pi^2t},\,\,\,
\int_{\mathbb{R}_+}\mathrm{d}s\, J_{1}(ts)s^2 f_2^1(x)\big|_{\mathbf{f}=\mathbf{f}_4}=\frac{4J_1^2(t/2)}{\pi^2t^3}-\frac{J_0(t)}{4\pi^2t},
\end{equation}
\begin{equation}
	\int_{\mathbb{R}_+}\mathrm{d}s\, J_{1}(ts)s^2 f_1^1(x)\big|_{\mathbf{f}=0}=\frac{J_1(t)}{2\pi^2t^2}-\frac{1}{4\pi^2t},\,\,\,
	\int_{\mathbb{R}_+}\mathrm{d}s\, J_{1}(ts)s^2 f_1^1(x)\big|_{\mathbf{f}=\mathbf{f}_4}=\frac{4J_1^2(t/2)}{\pi^2t^3}-\frac{1}{4\pi^2t}.
\end{equation}
The latter equalities must be understood in the sense of generalized functions, see \cite{Vladimirov-2002}. The third stage is the numerical integration of the expressions obtained. To do this, it is first necessary to replace the infinite limits of integration with suitable finite intervals and calculate the resulting integral over the finite interval using the composite trapezoid formula, see \cite{6,7}, using the Newton--Cotes formula. The method itself is quite simple in software implementation. For example, for the function $f(\cdot)\in C^2([a,b])$ when dividing the interval $[a,b]$ into $N$ parts, that is, in increments of $h =(b-a)/N$, the approximation of the integral is valid
\begin{equation}
	\int_a^b\mathrm{d}x\,f(x) = h\bigg(\frac{f(a)+f(b)}{2}+\sum_{i=1}^{N-1}f(x_i)\bigg)+R_n,
\end{equation}
where the argument values have the form $x_i = a + ih$. In this case, the absolute value of the error is $|R_n|$ with the order of accuracy $\mathcal{O}(h^2)$ is written out as
\begin{equation}
	|R_n| \leqslant h^2 \frac{b-a}{12}\times\max_{x\in[a,b]}\big|f''(x)\big|.
\end{equation}
Note that all calculations were performed in the ``Python'' programming environment using the ``NumPy'' computing library and the ``Numba'' computing acceleration package.

\section{Conclusion}
\label{v-s-6}
In this paper, a numerical analysis was performed of a number of integrals arising in the three-loop approximation of quantum action for the four-dimensional model with the quartic interaction. The integrals are given in Section \ref{v-s-3}, and their numerical values are given in Table\ref{table:1} in Section \ref{v-s-2}. At the same time, the values are presented in the table, including for the already known case, in order to verify the computational algorithm and additional comparison. Based on the values obtained, the third coefficients for the $\beta$-function and the corresponding anomalous dimensions were calculated, see Section \ref{v-s-2}. All the results are in agreement with the general theory.

Of particular interest is the option of calculating the function in the case of the cutoff regularization in the coordinate representation for the part of the renormalization constant corresponding to power singularities. As shown, the three-loop approximation allows us to study only the first two coefficients.

Another interesting example, from the point of view of applying the cutoff regularization, is the three-dimensional sextic model \cite{29-3}, in which it has now been possible to advance to four-loop coefficients, see \cite{Kh-25,v-61}. It is expected that in this model it may be possible to adapt the standard Bogoliubov--Parasyuk procedure \cite{33-rev7} in a very elegant way.

\vspace{2mm}
\noindent\textbf{Acknowledgments.} A.V. would like to thank N.V.Kharuk for useful comments.

\vspace{2mm}
\noindent\textbf{Data Availability Statement.} Data exchange is not applicable to this article because no data sets have been generated or analyzed during the current study.

\vspace{2mm}
\noindent\textbf{Code Availability Statement.} The corresponding code/software is not attached to the article.

\vspace{2mm}
\noindent\textbf{Conflict of interest statement.} The authors claim that there is no conflict of interest.


\begin{thebibliography}{99}
\bibitem{1}
L. D. Faddeev, A. A. Slavnov, \textit{Gauge Fields: An Introduction to Quantum Theory}, Frontiers in Physics \textbf{83}, Addison-Wesley, 1--236 (1991)
\bibitem{2}
M. E. Peskin, D. V. Schroeder, \textit{An Introduction to Quantum Field Theory}, Addison-Wesley, 1--868 (1995)

\bibitem{29-4}
H. Kleinert, V. Schulte-Frohlinde, \textit{Critical properties of $\phi^4$-theories}, World Scientific, Singapore, 1--512 (2001)
\bibitem{29-3}
A. N. Vasil'ev, \textit{The field theoretic renormalization group in critical behavior theory and stochastic dynamics}, Boca Raton: Chapman and Hall/CRC, 1--681 (2004)

\bibitem{19}
C. G. Bollini, J. J. Giambiagi, \textit{Dimensional Renormalization: The Number of Dimensions as a Regularizing Parameter}, Nuovo Cim. B, \textbf{12}, 20--26 (1972)
\bibitem{555}
G. 't Hooft, M. Veltman, \textit{Regularization and renormalization of gauge fields}, Nucl. Phys. B \textbf{44}, 189--213 (1972)

\bibitem{105}
D. I. Kazakov, \textit{Radiative Corrections, Divergences, Regularization, Renormalization, Renormalization Group and All That in Examples in Quantum Field Theory}, arXiv:0901.2208 [hep-ph] (2009)

\bibitem{v-2}
J. C. Collins, \textit{Scaling behavior of $\phi^4$ theory and dimensional regularization}, Phys. Rev. D, \textbf{10}, 1213--1218 (1974) doi:10.1103/PhysRevD.10.1213
\bibitem{v-3}
I. Jack und H. Osborn, \textit{Two–Loop Background Field Calculations for Arbitrary Background
Fields}, Nucl. Phys. B, \textbf{207}, 474--504 (1982) doi:10.1016/0550-3213(82)90212-7
\bibitem{v-15}
J. Zinn-Justin, \textit{Quantum Field Theory and Critical Phenomena}, Oxford Univ. Press, Oxford, 1--1054 (1989)

\bibitem{v-60}
T. Steudtner, A. E. Thomsen, \textit{General quartic $\beta$-function at three loops}, J. High Energ. Phys. \textbf{2024}, 163 (2024) doi:10.1007/JHEP10(2024)163

\bibitem{v-4}
E. Brezin, J. C. Le Guillou, J. Zinn-Justin, \textit{Addendum to Wilson's theory of critical phenomena and Callan--Symanzik equations in $4-\varepsilon$ dimensions}, Phys. Rev. D, \textbf{9}, 1121--1124 (1974) doi:10.1103/PhysRevD.9.1121
\bibitem{v-1}
A. A. Vladimirov, D. Kazakov, O. V. Tarasov, \textit{Calculation of critical exponents by Quantum Field Theory Methods}, Sov. Phys. JETP, \textbf{50}, 521--526 (1979)
\bibitem{v-5}
M. V. Kompaniets, E. Panzer, \textit{Renormalization group functions of $\phi^4$ theory in the $\mathrm{MS}$-scheme to six loops}, PoS, \textbf{260}, LL2016, 038 (2016) doi:10.22323/1.260.0038
\bibitem{v-55}
A. Bednyakov, A. Pikelner, \textit{Six-loop beta functions in general scalar theory}, J. High Energ. Phys. \textbf{2021}, 233 (2021) doi:10.1007/JHEP04(2021)233

\bibitem{chi-0}
A. Brizola, O. Battistel, M. Sampaio, M. C. Nemes, \textit{Implicit Regularisation Technique: Calculation of the Two-loop $\phi^4_4$-theory $\beta$-function}, Mod. Phys. Lett. A, \textbf{14}, 1509--1518 (1999)

\bibitem{v-11}
M. Pernici, M. Raciti, \textit{Wilsonian flow and mass-independent renormalization}, Nucl. Phys. B, \textbf{531}, 560--592 (1998) doi:10.1016/S0550-3213(98)80007-2
\bibitem{v-10}
T. R. Morris, J. F. Tighe, \textit{Convergence of derivative expansions of the renormalization group}, JHEP \textbf{08}, 007 (1999) doi:10.1088/1126-6708/1999/08/007
\bibitem{v-12}
P. Kopietz, \textit{Two-loop $\beta$-function from the exact renormalization group}, Nucl. Phys. B, \textbf{595}, 493--518 (2001) doi:10.1016/S0550-3213(00)00680-5

\bibitem{v-13}
K. Wilson, J. Kogut, \textit{The renormalization group and the epsilon expansion}, Phys.
Rep. \textit{12}, 75--199 (1974) doi:10.1016/0370-1573(74)90023-4
\bibitem{v-14}
J. Polchinski, \textit{Renormalization and effective lagrangians}, Nucl. Phys. B, \textbf{231}, 269--295 (1984) doi:10.1016/0550-3213(84)90287-6




\bibitem{3}
A. V. Ivanov, \textit{Three-loop renormalization of the quantum action for a four-dimensional scalar model with quartic interaction with the usage of the background field method and a cutoff regularization}, Nucl. Phys. B, \textbf{1006}, 116647 (2024), doi:10.1016/j.nuclphysb.2024.116647, arXiv:2402.14549
\bibitem{4}
A. V. Ivanov, \textit{Explicit Cutoff Regularization in Coordinate Representation}, 2022 J. Phys. A: Math. Theor. \textbf{55}, 495401, arXiv:2209.01783, doi:10.1088/1751-8121/aca8dc


\bibitem{5}
A. V. Ivanov, \textit{An applicability condition of a cutoff regularization in the coordinate representation}, Funct Anal Its Appl \textbf{59}, 1--10 (2025) arXiv:2403.09218, doi:10.1134/S123456782501001X

\bibitem{Vladimirov-2002}
V. S. Vladimirov, \textit{Methods of the theory of generalized functions}, London,
CRC Press, 1--328 (2002)
\bibitem{6}
N. N. Kalitkin, \textit{Numerical Methods, 2nd edition}, BHV, St. Petersburg, 1--592 (2011) 
\bibitem{7}
A. A. Samarsky, A. V. Gulin, \textit{Numerical Methods}, Moscow: Nauka, 1--432 (1989) 


\bibitem{Kh-25}
N. V. Kharuk, \textit{Four-loop renormalization with a cutoff in a sextic model}, 2025 J. Phys. A: Math. Theor. \textbf{58}, 395401 (2025) arXiv:2504.07688 doi:10.1088/1751-8121/ae0798
\bibitem{v-61}
J. Gaite, \textit{Renormalization group for effective field theories: Cutoff schemes and universality}, Nucl. Phys. B, \textbf{1019}, 117109 (2025) doi:10.1016/j.nuclphysb.2025.117109

\bibitem{33-rev7}
O. I. Zavialov, \textit{Bogolyubov's $\mathcal{R}$-operation and the Bogolyubov--Parasyuk theorem}, Russian Math. Surveys, \textbf{49}:5, 67--76 (1994) doi:10.1070/RM1994v049n05ABEH002426



\end{thebibliography}
\end{document}